\documentclass[preprintnumbers]{revtex4}
\usepackage{eurosym}
\usepackage{amsmath}
\usepackage{amssymb}
\usepackage{graphicx}
\usepackage{color}
\usepackage{epstopdf}
\usepackage{epstopdf}
\DeclareUnicodeCharacter{2212}{-}
\usepackage{bm}

\begin{document}

\date{\today }
\title{Electric-field-tuned binding energies of trions in silicene, germanene, and stanene monolayers }
%Electric-field tunable binding energies of trions in silicene, germanene and stanene monolayers.}
\author{Roman Ya. Kezerashvili$^{1,2}$, Shalva M. Tsiklauri,$^{3}$ and Anastasia Spiridonova$^{1}$}
\affiliation{\mbox{$^{1}$New York City College of Technology, The City University of New York, USA} \\
$^{2}$The Graduate School and University Center, The City University of New
York, USA\\
$^{3}$Borough of Manhattan Community College, The City University of New
York, USA}

\begin{abstract}
We predict the formation of intravalley controllable trions in  buckled two-dimensional (2D) materials such as silicene, germanene, and stanene
monolayers in an external electric field. Performing a study within the framework of a nonrelativistic potential model using
the method of hyperspherical harmonics (HH), the three-body Schr\"{o}dinger equation is solved with the Rytova-Keldysh potential by expanding
the wave functions of a trion in terms of the
HH. Then, we numerically solve a resultant system of coupled differential equations. The ground state energies of intravalley trions controlled by the external electric field are presented. The dependencies of the binding energy (BE) of trions in silicene, germanene, and stanene as a function of the electric field are shown to be qualitatively similar. BEs of trions formed by $A$ and $B$ excitons have a non-negligible difference that increases slightly as the electric field increases. We demonstrate that trion BEs can be controlled by the external electric field.

\end{abstract}

\maketitle

\section{Introduction}
The prediction of the existence of trions \cite{Lampert58}
consisting of an exciton and an electron or a hole, known as negatively or
positively charged excitons (X$^{\mp }$), gave rise to many theoretical and
experimental studies of trions in bulk materials, quantum-well systems, and two-dimensional (2D) materials.
Atomically thin transition-metal dichalcogenides (TMDCs) are a class of 2D materials that have
remarkable optical and electronic properties \cite{Kormanyos, RMP}.
Since the observation of trions in two-dimensional MoS$_{2}$
monolayers \cite{MoS23Heinz} in 2013, trions have been the subject of intense studies, both experimentally and
theoretically, in TMDCs monolayer. In the past decade, different experimental
groups have observed and reported the signature of a trion in TMDCs monolayers.

Theoretical studies of trions have integrated a wide variety of techniques and carried them out to calculate the binding energies (BEs) of
excitonic complexes in monolayer TMDCs (see the reviews \cite%
{NScRev2015,BekReichman,Durnev2018,Kezerashvili2019,Suris2022}). Results for BEs of trions in TMDC monolayers yielded
impressively accurate results consistent with experimental data.
In the framework of the few-body physics approaches such as the hyperspherical harmonic method and three-body Faddeev equations in configuration or momentum spaces
trions in TMDC are investigated in \cite{KezFew2017, FilikhinKezPL2018,Frederico2023,Keztsik2024}.

Another category of 2D semiconductors is the buckled 2D allotropes of silicon, germanium, and tin, known as silicene, germanene, and stanene, and collectively referred to as Xenes \cite{Molle2017, Zheng2020}. Experimental studies revealed one of the most crucial differences between Xenes and graphene and TMDC that Xenes monolayer is not a perfectly flat sheet, but instead it is slightly buckled \cite{Matthes2013a,Molle2017}. As a result, this unique structure of Xenes makes them sensitive to the external electric field applied perpendicular to the monolayer, allowing the band gap to be opened and controlled. The tunable
band gap of Xenes gives researchers, among other things, extraordinary in situ control over
binding energies and optical properties of excitons in these materials.

In contrast to TMDCs, there is no extensive research on excitonic complexes in Xenes monolayers. The reason is that the synthesis of Xenes monolayers has not been as successful and extensive, as for example, TMDCs monolayers because Xenes monolayers are unstable in the air \cite{Acun2013,Tao2015}. In contrast to graphene, silicene and other Xenes monolayers do not occur in nature. Nevertheless, silicene nanoribbons were experimentally synthesized on a metal substrate \cite{Aufray2010, Padova2010}.
This opened the way for silicene, germanene, and stanene monolayers to be transferred on metal such as Au \cite{Drummond2012,Ezawa2,Davila2014,Mannix2017,Grazianetti2020} and substrates such as MoS$_2$, Ir, ZrB$_2$ \cite{Mannix2017,Sante2019} and hexagonal boron nitride (hBN) \cite{Li2013, Khan2017} and synthesized as freestanding monolayers \cite{Saxena2016}. Working with a metallic substrate is easier. For example, silicene grown on Ag (111) \cite{Ezawa2012sil,Tao2015} and germanene synthesized by dry deposition on Au (111) surface \cite{Davila2014} have been thoroughly investigated. However, depositing Xene on a metal leads to a significant alteration of properties of the Xene monolayer. Depending on a substrate, the properties of Xenes monolayers vary; see Refs. \cite{Molle2017,Li2013} for the list of Xenes properties on different substrates.
In contrast to deposition on a metal, depositing Xenes on hBN is harder. However, Xene deposited on hBN preserves its properties because Xene and hBN weakly interact \cite{Li2013}.  Xenes monolayers deposited on hBN present a particular interest for studying magnetoexcitons in monolayers and vdWHs \cite{KezSpir2022Xenes}.

Xenes optical and magneto-optical properties have been addressed in Refs. \cite{Bechstedt2012,Stille2012,Matthes2013a,Fadaie2016} and \cite{Muoi2020,Chowdhury2016}, respectively. Different physical phenomena, such as the Hall effect \cite{Ezawa2012sil}, the valley-locked spin-dependent Seebeck effect \cite{Zhai2018}, the anomalous quantum Hall effect \cite{Ezawa2}, the quantum spin Hall effect \cite{Zhao2020}, and the Landau levels \cite{Ezawa2012sil, Tsaran2014,Muoi2020}  are studied because of their essential role in applications of Xenes monolayers in nanodevices and quantum  devices \cite{Chen2018,  Lyu2019,Glavin2020, Tao2020, Zhao2020}.

Because of the band inversion, these honeycomb materials are also topological insulators \cite{Bechstedt2021,Liu2011,Ezawa2015,Matthes2016,Matusalem2017,Yu2017}. The existence of an excitonic insulator phase in silicene, germanene, and stanene was first studied in \cite{BBKPL2019,BBKPRB2019} in the framework of the effective-mass approximation.
The influence of the screening, band dispersion, and external electric field on transitions in Xenes between excitonic, topological, and trivial insulator phases was investigated in \cite{Pulci2023}.

Currently, there is a shortage of research on exciton complexes in Xenes. In particular, there has not been a study of the formation of three quasiparticle states trions in Xenes monolayers. %and their binding energies.
In this paper, we address this gap and focus on a theoretical investigation of trions in Xenes within the method of hyperspherical harmonics (HH).

The
paper is organized as follows. In Sec. \ref{Theory}, we present
nonrelativistic potential model for a system of three interacting electrons and holes,
and we employ the three-body Schr\"{o}dinger equation in the effective-mass approximation. Following \cite{Keztsik2024} in the framework of the HH method, the Schr\"{o}dinger
equation with the Rytova-Keldysh potential \cite{Ritova,Keldysh} is reduced to a system
of coupled differential equations for
hyperradial functions. A numerical solution of this system provides the BE and wave function for trions in Xenes. In Sec. \ref{Trions}, we discuss intravalley $X^{-}$ and $X^{+}$  trions in Xenes monolayers and present results of calculations of controllable ground state energies of intravalley trions by an external electric field. Here, we analyze the dependence of the BEs and a probability distribution of three bound particles on the external electric field. Conclusions follow in Sec. \ref{conc}.

\section{%Theoretical model \\
Effective-mass approach for trions in
buckled 2D materials}
\label{Theory}

The starting point of the discussion needs to be the framework of the low-energy model for excitons in Xenes monolayer and heterostructure when the external electric field perpendicular to the structure is present. The detailed overview of the low-energy model for excitons in the buckled 2D material can be found in Refs. \citep{BBKPRB2019,KezSpir2022Xenes}. The most stable form of Xenes monolayers has the honeycomb structure where sublattices A and B are offset with respect with each other. The offset is denoted by $d_0$ and is called the buckling constant. When there is no external electric field, Xenes monolayers resemble graphene in the vicinity of the $K/K'$ points. However, the application of the perpendicular electric field leads to on-site potential difference between sublattices, that results in the variable band gap at the $K/K'$ points that changes electron and hole effective masses. Considering the electric-field-dependent band gap at $p=0$ \cite{Tabert}, the effective mass can be written as:

\begin{equation}
m=\frac{\lvert \xi \sigma \Delta _{gap}-ed_{0}E_{\perp }\rvert }{v_{F}^{2}},
\label{eq:effmassEz}
\end{equation}%
 where $v_F$ is the Fermi velocity, $\xi ,\sigma =\pm 1$ are the
valley and spin indices, respectively, 2$\Delta_{gap}$ is the intrinsic band gap, $E_{\perp }$ is the external electric field in $z$-direction. When $\xi =-\sigma $, the gap is
large, since conduction and valence bands are the furthest from the
Fermi energy level, and the electron and hole form $A$ excitons. When $\xi =\sigma $, the gap is small and the electron and hole form $B$ exciton.
 According to Eq. (\ref%
{eq:effmassEz}) the effective mass of the electron or hole depends on the band
gap, Fermi velocity, buckling constant and is a function of $E_{\perp }$.

The promotion of an electron from the filled valence
band to the empty conduction band leaves an empty electron
state in the valence band. The description of such
a many-body system can be reduced to the two-particle
problem of a negatively charged conduction electron interacting
with a positively charged valence hole that forms an exciton or other excitonic complexes such as charged excitons or trions. The trions are formed when an
exciton binds another electron or hole to form a negatively or positively charged three-particle system: $X^{-}$ or $X^{+}$, respectively. A description of the properties of trions requires a solution of a three-particle problem. In buckled two-dimensional monolayers, the resulting trions are considered as Wannier-Mott trions since
the correlation between an electron and a hole extends
over many lattice periods. The representation of the electron-hole pair bound in a Wannier-Mott exciton shows the strong spatial correlation of these two constituents. Therefore, we are assuming that the interaction of the exciton with the third particle (electron or hole) leads to a large-radius-type system.

We follow the approach in which one
assumes that the electron and hole bands are isotropic and parabolic, which
is a good approximation for the low-energy spectrum of 2D materials. This
form of the Hamiltonian implies that both the electron and hole single
particle states form a single parabolic band. The corresponding eigenproblem
equation reduces to the Schr\"{o}dinger equation in the effective mass
approximation. This approach is commonly used in the literature to describe
excitons and trions in 2D materials. See, for example, Refs.~%
\cite{Kormanyos,BekReichman,Kezerashvili2019,Keztsik2024,Berkelbach2013,Fogler2014,Berman2016,Wang2018,Berman2017}.
Below, we follow the effective-mass approximation.

To obtain the eigenfunctions and eigenenergies of a 2D trion in
Xenes when the electric field is perpendicular to the Xenes monolayer, we write
the Schr\"{o}dinger equation for an interacting three-particle electron-hole
system. Because we are considering the varying electric field $E_{\perp }$,
which is directed along the $z$-axis, the corresponding term in the 2D Schr\"{o}%
dinger equation vanishes. However, the effect of the electric field action
is present through the effective mass as follows from Eq. (\ref{eq:effmassEz}). Thus, one can write the 2D Schr\"{o}%
dinger equation for the interacting three-particle electron-hole system
within the effective-mass approximation in the following form

\begin{equation}
\left[ -\frac{\hslash ^{2}}{2}\overset{3}{\underset{i=1}{\sum }}\frac{1}{%
m_{i}}\nabla _{i}^{2}+\overset{3}{\underset{i<j}{\sum }}V_{ij}(\left\vert
\mathbf{r}_{i}-\mathbf{r}_{j}\right\vert )\right] \Psi (\mathbf{r}_{1},%
\mathbf{r}_{2},\mathbf{r}_{3})=E\Psi (\mathbf{r}_{1},\mathbf{r}_{2},\mathbf{r%
}_{3}),  \label{Trion}
\end{equation}%
where $m_{i}$ is the effective mass of the electron or hole defined by Eq. (\ref{eq:effmassEz}), and $\mathbf{r}_{i}$ is the $i$th
particle position. We assume only two types of charge
carriers: an electron and hole with the corresponding effective masses. In
Eq. ~(\ref{Trion}), $V_{ij}(\left\vert \mathbf{r}_{i}-\mathbf{r}%
_{j}\right\vert )$ is the interaction potential between $q_{i\text{ }}$ and $%
q_{j}$ charges in a 2D material that was first derived in Ref. \cite{Ritova}
and was independently obtained by Keldysh \cite{Keldysh}. We refer to it as
the Rytova-Keldysh (RK) potential. The Rytova-Keldysh potential describes
the Coulomb interaction screened by the polarization of the electron
orbitals in the 2D lattice and has the following form

\begin{equation}
V_{ij}(r)=\frac{\pi kq_{i}q_{j}}{2\kappa \rho _{0}}\left[ H_{0}(\frac{r}{%
\rho _{0}})-Y_{0}(\frac{r}{\rho _{0}})\right],   \label{Keldysh}
\end{equation}%
where $r=\left\vert \mathbf{r}_{i}-\mathbf{r}%
_{j}\right\vert $ is the relative coordinate between two charge carriers $q_{i}$ and $q_{j}$. In
Eq.~(\ref{Keldysh}) $k=9\times 10^{9}$ N$\cdot $m$^{2}$/C$^{2}$, $\kappa $
is the dielectric constant of the environment that is defined as $\kappa
=(\varepsilon _{1}+\varepsilon _{2})/2$, where $\varepsilon _{1}$ and $%
\varepsilon _{2}$ are the dielectric constants of two materials that the
Xenes layer is surrounded by%either below and above the monolayer
, $\rho _{0\text{ }}$ is the screening length, which sets the boundary between two
different behaviors of the potential due to a nonlocal macroscopic
screening, and $H_{0}(\frac{r}{\rho _{0}})$ and $Y_{0}(\frac{r}{\rho _{0}})$
are the Struve function and Bessel function of the second kind,
respectively. The screening length $\rho _{0}$ can be written as $\rho _{0}=(2\pi \chi _{2D})/(\kappa )$  \cite%
{Berkelbach2013}, where $\chi _{2D}$
is the 2D polarizability, which in turn is given by $\chi
_{2D}=l\varepsilon /4\pi $~\cite{Keldysh}, where $\varepsilon $ is the bulk dielectric constant
of the Xene monolayer.
\iffalse
For large distances $\ r>>\rho _{0\text{ }}$the
potential has the three-dimensional bare Coulomb tail $V_{ij}(r\mathbf{)=}%
\frac{kq_{i}q_{j}}{\epsilon r}$, while at very small distances, smaller
than the screening length $r<<\rho _{0\text{ }}$, it becomes a logarithmic
potential like a potential of a point charge in two dimensions: $V_{ij}(r%
\mathbf{)=}\frac{kq_{i}q_{j}}{\epsilon \rho _{0\text{ }}}\left[ \ln \left(
\frac{r}{2\rho _{0}}\right) +\gamma \right] $, where $\gamma $ is the Euler
constant. Therefore, the potential (\ref{Keldysh}) becomes the standard
bare Coulomb potential at $r>>\rho _{0\text{ }}$ and diverges
logarithmically at $r<<\rho _{0\text{ }}.$ A crossover between these two
regimes takes place around distance $\rho _{0}$. Thus, at small distances
between charge carriers the short-range interaction strength decreases,
while the long-range interaction strength is unaffected and is the bare
three-dimensional Coulomb potential.
\fi
It is worth noting that in Ref. \cite%
{Rubio} a very good approximation to the RK potential
that is simpler to use, fairly precise in both limits, and remarkably
accurate for all distances was introduced.

To obtain a solution of the Schr\"{o}dinger equation (\ref{Trion})
for the negatively and positively charged trions, we use the method of
hyperspherical harmonics (HH) \cite{Avery}. The main idea of this method is
the expansion of the wave function of the trion in terms of \ HH that are
the eigenfunctions of the angular part of the Laplace operator in the
four-dimensional (4D) space. As the first step, let us separate the
center-of-mass (c.m.) and the relative motion of three particles and introduce
sets of mass-scaled Jacobi coordinates \cite{Avery,JibutiSh}. There are three equivalent sets of Jacobi coordinates and there is the orthogonal transformation between these
sets \cite{FM,FilKez2024}. For three non-identical particles that have different masses the mass-scaled Jacobi coordinates %read In Xenes the mass of electrons and holes are equal. One can write Jacobi coordinates
for the partition $i$ read as follows \cite{Avery,JibutiSh,FM}:
\begin{eqnarray}
\mathbf{x}_{i} &=&\sqrt{\frac{m_{j}m_{k}}{(m_{j}+m_{k})\mu }}(\mathbf{r}_{j}-%
\mathbf{r}_{k}),  \notag \\
\mathbf{y}_{i} &=&\sqrt{\frac{m_{i}\left( m_{j}+m_{k}\right) }{%
(m_{i}+m_{j}+m_{k})\mu }}\left( \frac{m_{j}\mathbf{r}_{j}+m_{k}\mathbf{r}_{k}%
}{m_{j}+m_{k}}-\mathbf{r}_{i}\right) ,\text{ \ }i\neq j\neq k=1,2,3,
\label{Jacobi3}
\end{eqnarray}%
where
\begin{equation}
\mu =\sqrt{\frac{m_{i}m_{j}m_{k}}{m_{i}+m_{j}+m_{k}}}\ \   \label{EffecMass}
\end{equation}%
is the three-particle effective mass.
% and $\mathbf{R}$ is the three-particle c.m. coordinate.
In Eqs. (\ref{Jacobi3}) the subscripts $i$, $j$, and $k$ are a cyclic
permutation of the particle numbers.

The transformation (\ref{Jacobi3}%
) allows the separation of c.m. and relative motions of three particles
with Hamiltonian (\ref{Trion}), and the Schr\"{o}dinger equation for the
relative motion of the three-body system reads
\begin{equation}
\left[ -\frac{\hbar ^{2}}{2\mu }(\nabla _{x_{i}}^{2}+\nabla
_{y_{i}}^{2})+V(x_{1})+V(x_{2})+V(x_{3})\right] \Psi (\mathbf{x}_{i},\mathbf{%
y}_{i})=E\Psi (\mathbf{x}_{i},\mathbf{y}_{i}).\ \   \label{Relative3}
\end{equation}%
In Eq. (\ref{Relative3}), $V(x_{i})$ is the interaction potential between two
particles %separated by distances
at the relative distance $x_{1}$, $x_{2}$, and $x_{3}$, respectively, where $%
x_{i}$ is the modulus of the Jacobi vector $\mathbf{x}_{{i}}$ (\ref{Jacobi3}%
), and (\ref{Relative3}) is written for any of the set $i=1,2,3$ of the Jacobi
coordinates (\ref{Jacobi3}). The orthogonal transformation between three
equivalent sets of the Jacobi coordinates %(see, for example, \cite%{FM,FilKez2024})
simplifies calculations of matrix elements involving $%
V(x_{i})$ potentials.

This method is presented in detail in Ref. \cite{Keztsik2024} and is briefly outlined here. We introduce in the 4D space the hyperradius $\rho =%
\sqrt{x_{i}^{2}+y_{i}^{2}}$ and a set of three angles $\Omega _{i}\equiv
(\alpha _{i},\varphi _{x_{i}},\varphi _{y_{i}}),$ where $\varphi _{x_{i}}$
and $\varphi _{y_{i}}$ are the polar angles for the Jacobi vectors $\mathbf{x%
}_{i}$ and $\mathbf{y}_{j},$ respectively, and $\alpha _{i}$ is an angle
defined as $x_{i}=\rho \cos \alpha _{i},$ $y_{i}=\rho \sin \alpha _{i}$.
Next, we rewrite the Schr\"{o}dinger equation (\ref{Relative3}) for the trion
using hyperspherical coordinates in the 4D configuration space \cite{Keztsik2024}. This
transformation allows us to reduce the solution of the problem for the three
particles in the 2D configuration space to the motion of one particle in the 4D
configuration space. Then we introduce the hyperspherical harmonics $\Phi _{K\lambda }(\Omega )$ in the 4D configuration space, which are the eigenfunctions of the angular part of the generalized Laplace
operator $\widehat{K}^{2}(\Omega _{i})$  in the 4D configuration space $\widehat{K}^{2}(\Omega _{i})\Phi _{K\lambda }(\Omega )=K(K+2)\Phi
_{K\lambda }(\Omega )$ \cite{ Avery}, where $K$ is a grand angular momentum. Here we are using the short-hand notation $\lambda \equiv $ $%
\{l_{x},l_{y},L,M\},$ where $L$ is the total orbital angular momentum of the
trion, $M$ is its projection, and the grand angular momentum $K=2n+l_{x}+l_{y}$, $l_{x}$, where $l_{x}$ and $l_{y}$ are the angular momentum corresponding to
$\mathbf{x}$ and $\mathbf{y}$ Jacobi coordinates, respectively, and $n$ $\geqslant 0$ is an
integer number.

The functions $\Phi _{K\lambda }(\Omega )$ present a
complete set of orthonormal basis, and one
can expand the wave function of the trion $\Psi (\rho ,\Omega _{i})$ in
terms of the HH $\Phi _{K\lambda }(\Omega )$ as
\begin{equation}
\Psi (\rho ,\Omega _{i})=\rho ^{-3/2}\sum_{_{K\lambda }}u_{K\lambda }(\rho
)\Phi _{K\lambda }(\Omega _{i}).\   \label{ExpanTrion}
\end{equation}%
In Eq. (\ref{ExpanTrion}), $u_{K\lambda }(\rho )$ are the hyperradial
functions,
and by substituting (\ref{ExpanTrion}) into the Schr\"{o}dinger equation written in the hyperspherical coordinates \cite{ Keztsik2024}, one
can separate the radial and angular variables and get a set of coupled
differential equations for the hyperradial functions $u_{K\lambda }(\rho )$:

\begin{equation}
\left[ \frac{d^{2}}{d\rho ^{2}}-\frac{(K+1)^{2}-1/4}{\rho ^{2}}+\kappa ^{2}%
\right] u_{K\lambda }(\rho )=\frac{2\mu }{\hbar ^{2}}\sum_{_{K^{^{\prime
}}\lambda ^{^{\prime }}}}\mathcal{W}_{K\lambda K^{^{\prime }}\lambda
^{^{\prime }}}(\rho )u_{K^{^{\prime }}\lambda ^{^{\prime }}}(\rho ).
\label{TrionGeneral}
\end{equation}%
In Eq. (\ref{TrionGeneral}) $\kappa ^{2}=2mB/\hslash ^{2},$ where $B$ is
trion BE. The coupling effective potential energy $\mathcal{W}%
_{K\lambda K^{^{\prime }}\lambda ^{^{\prime }}}(\rho )$ is
\begin{equation}
\mathcal{W}_{K\lambda K^{^{\prime }}\lambda ^{^{\prime }}}(\rho )=\int \Phi
_{K\lambda }^{\ast }(\Omega _{i})\sum_{i<j}^{3}V_{ij}(\left\vert \mathbf{r}%
_{i}-\mathbf{r}_{j}\right\vert )\Phi _{K^{^{\prime }}\lambda ^{^{\prime
}}}(\Omega _{i})d\Omega _{i}.  \label{W3general}
\end{equation}

The coupling effective interaction (\ref{W3general}) is defined via
the RK potential~(\ref{Keldysh}). Substituting (\ref{Keldysh}) into Eq. (\ref%
{W3general}), one obtains the matrix elements of the effective potential energies. %(\ref{K_avarage}),
The method of calculations of the effective potential energies is given in \cite{Keztsik2024}.
Calculations of matrix elements $\mathcal{W}_{K\lambda K^{^{\prime }}\lambda ^{^{\prime }}}(\rho )$ of the two-body $V_{ij}(\left\vert \mathbf{r}%
_{i}-\mathbf{r}_{j}\right\vert )$ interactions
in the hyperspherical harmonics expansion method
for a three-body system are greatly simplified by using the HH basis states appropriate for the partition corresponding
to the interacting pair. Using the matrix elements $\mathcal{W}_{K\lambda K^{^{\prime }}\lambda ^{^{\prime }}}(\rho )$ in Eq. (\ref{TrionGeneral}), one can solve the system of coupled differential equations numerically. Results of numerical solutions of (\ref{W3general}) for trions in Xenes are presented in the next section.

\section{Intravalley trions in Xenes}
\label{Trions}

\begin{figure}[tbh]
\centering
\includegraphics[width=5cm]{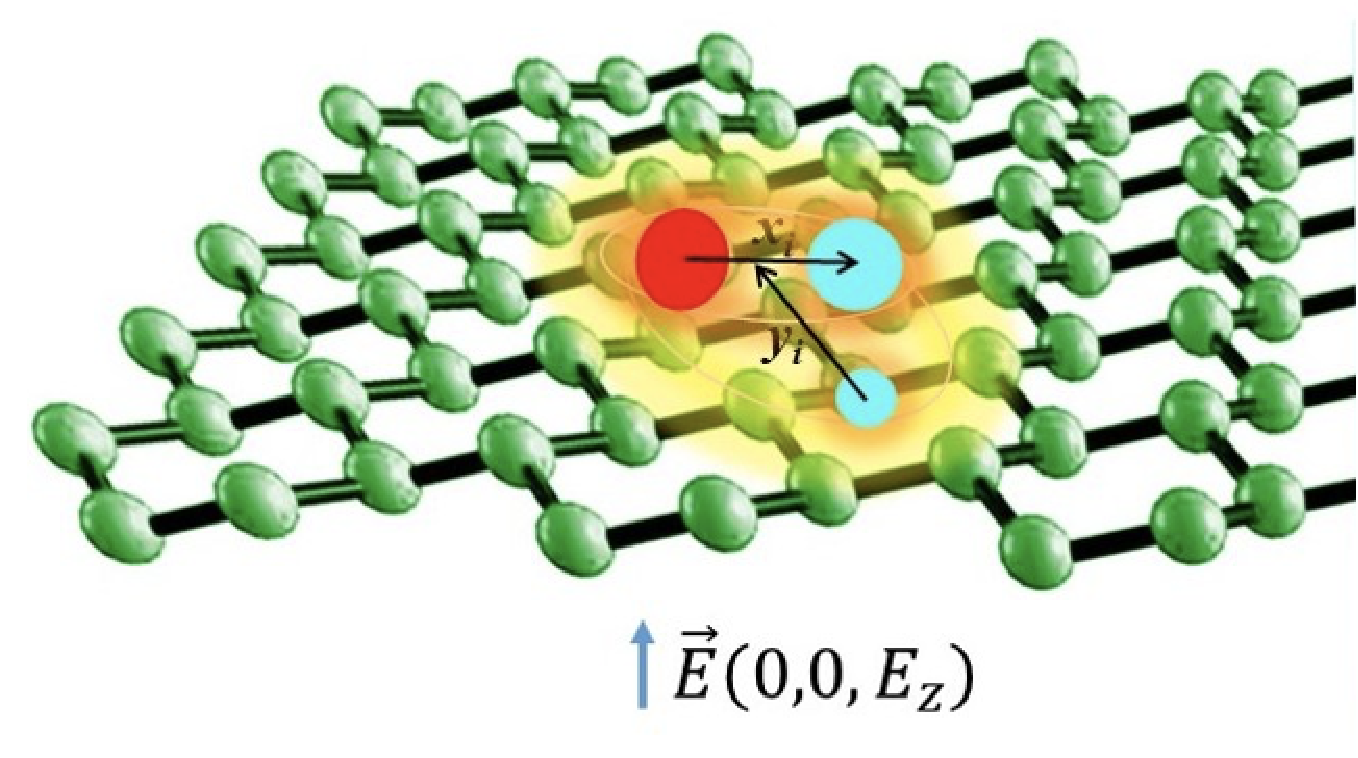}
\caption{(Color online) Schematic representation of the X$^{-}$ trion in the external electric field perpendicular to the freestanding Xenes  layer. $x_{i}$
and $y_{i}$ are Jacobi coordinates for the partition $i$. }
\label{TrionMon}
\end{figure}

We apply the present theoretical approach for calculations
of the trion BEs in the following freestanding Xenes
monolayers: silicene (Si), germanene (Ge), and stanene (Sn) in the external electric field $E_{z}=E_{\perp }$ perpendicular to the Xene layer, as shown in Fig. \ref{TrionMon}. The form
of the trion wave function (\ref{ExpanTrion}) is the most general, not restricted
to any particular mass ratio of electrons and holes and
describes the three-particle relative motion. The splitting of the conduction and valence bands in Xenes due to spin–orbit coupling at non-zero electric fields leads to the formation of $A$ and $B$ excitons in the larger or smaller band gaps, with corresponding larger or smaller masses of the electron and hole. Two of the three particles constituting a positive or a negative trion in the Xene monolayer have the same masses due to the equity of the mass of electron and hole that form an exciton. However, they are not identical because they have different charges. Below we consider trions formed by a singlet bright  $A$ or $B$ exciton and an electron (X$^{-}$) or hole (X$^{+}$).

\begin{figure}[tbh]
\centering
\includegraphics[width=10.0cm]{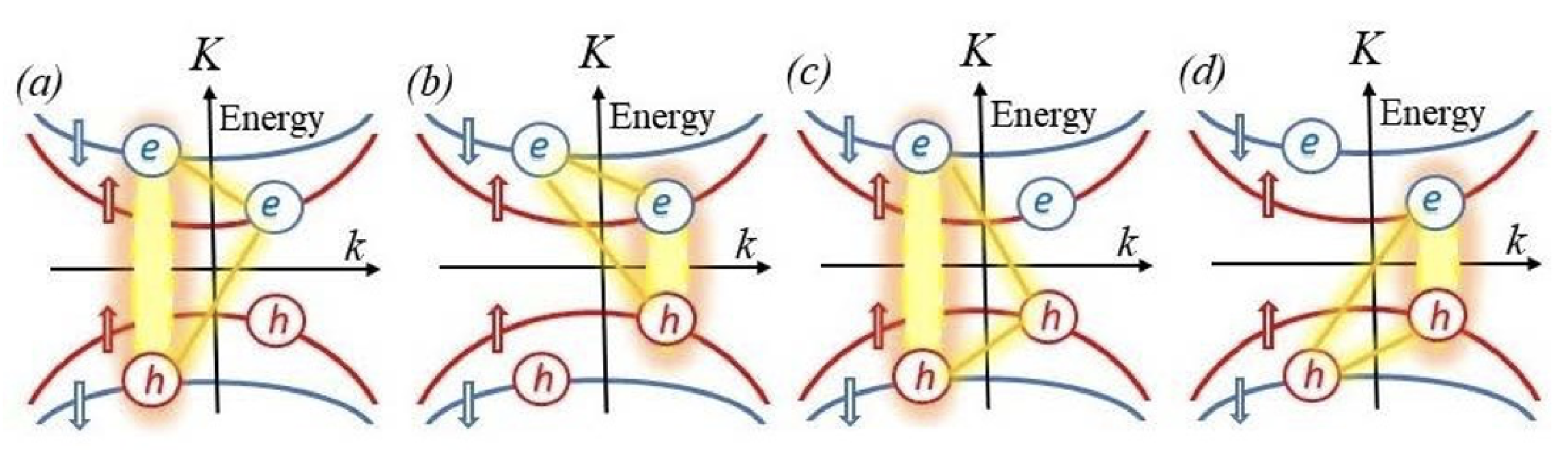}
\caption{(Color online) Schematic representation of the low-energy band
structure for 2D Xenes material and formation of the intravalley $X^{-}$ and
$X^{+}$ trions. Panels ($a$), ($b$) and ($c$), ($d$) represent the intravalley $X^{-}$ and $X^{+}$, respectively.}
\label{KvalleyTrion}
\end{figure}

%\subsection{Intravalley trions}
%\label{IntraInter}
In 2D monolayer of Xenes intravalley and intervalley trions can be formed. Here we consider only intravalley trions. An interaction of bright the $A$ or $%
B$ exciton with another charged carrier in the same valley, which can
either be an electron or a hole, forms $X^{-}$ or $X^{+}$ intravalley
trion. Schematics in Fig. \ref{KvalleyTrion} show the possible formations of the $X^{-}$
(Figs. \ref{KvalleyTrion}$a$ and 2$b$) and $X^{+}$ (Figs. \ref{KvalleyTrion}$c$ and 2$d$). As follows from Eq. (\ref{eq:effmassEz}),  the electron and hole
effective masses in $A$ exciton are the same. The $B$ excitons are composed of electron and hole which also have equal masses but smaller than masses of the electron and hole in $A$ excitons. We denoted these masses as $m_{A}$ and $m_{B}$, respectively. The intravalley $X^{-}$ and $X^{+}$ trions have the same two particles' masses. As follows from Eq. (\ref{EffecMass}), the effective masses of the $X^{-}$
and $X^{+}$ trions formed by $A$ or $B$ exciton are  $\mu _{A}=\sqrt{\frac{%
m_{B}m_{A}^{2}}{m_{B}+2m_{A}}}$ and  $\mu _{B}=\sqrt{\frac{m_{A}m_{B}^{2}}{%
m_{A}+2m_{B}}}$, respectively \cite{Keztsik2024}. Because $m_{A}>m_{B}$, it follows that $\mu _{A}>\mu _{B}$. Therefore, the effective mass of the $X^{-}$ and $X^{+}$ trions formed by the $A$ exciton is larger than mass of the $X^{-}$ and $X^{+}$ trions formed by the $B
$ exciton. Due to the proportionality of the BE of trions to the three-particle effective mass $\mu $  \cite{Keztsik2024}, a BE of $X^{-}$ and $X^{+}$ trions formed by the $A$
exciton is larger than BE of trions formed by the $B$ exciton. In the ground state, both intravalley trions formed by charge carriers from the same valley are spin-singlet trions. The intervalley
trions can exist in either the singlet or triplet states. They are not considered in this work.
%The intravalley X− trion in the spin-singlet state, with both electrons and a hole in the same valley, is a bright spin singlet trion have a total spin 1/2 and are a spin-singlet trions.
% Because the BE of trions proportional to the effective mass $\mu $  \cite{Keztsik2024} a BE of $X^{-}$ and $X^{+}$ trions formed by $A$ exciton is larger than by the $B$ exciton

%\subsection{\protect\bigskip Results of calculations}
%\label{ResCal}

The intravalley $X^{-}$ and $X^{+}$ trions in Xenes monolayers have two particles (the electron and hole) that have the same masses, and the third particle (the electron or hole) has a different mass, as seen in Figs. \ref{KvalleyTrion}$a-2d$.
Therefore, we have to deal with three non-identical particles because two particles with the same masses have different charges.

In calculations of the BEs of trions in Xenes monolayers, we use the RK potential. We solve the system of coupled differential equations (\ref{TrionGeneral}) for the
hyperradial functions $u_{K\lambda }^{L}(\rho )$ numerically. By
solving the system of equations (\ref{TrionGeneral}), one finds the binding
energy as well as the corresponding hyperradial functions. The latter allows
one to construct the wave function (\ref{ExpanTrion}). Numerical solution of the coupled differential equations requires the control of the convergence of the BEs for trions with respect to the grand angular momentum $%
K $ for each value of the external electric field. The relative convergence of the BE is checked as $\Delta B/B=\left[ B(K+2)-B(K)\right] /B(K)$, where $B(K)$ is the
BE for the given $K$. The analysis of the results for the BEs at different values of the electric field shows that the reasonable convergence is reached for $%
K_{\max }$ = 14, so we limit our considerations to this value.

\begin{figure}[tbh]
\centering
\includegraphics[width=5 cm]{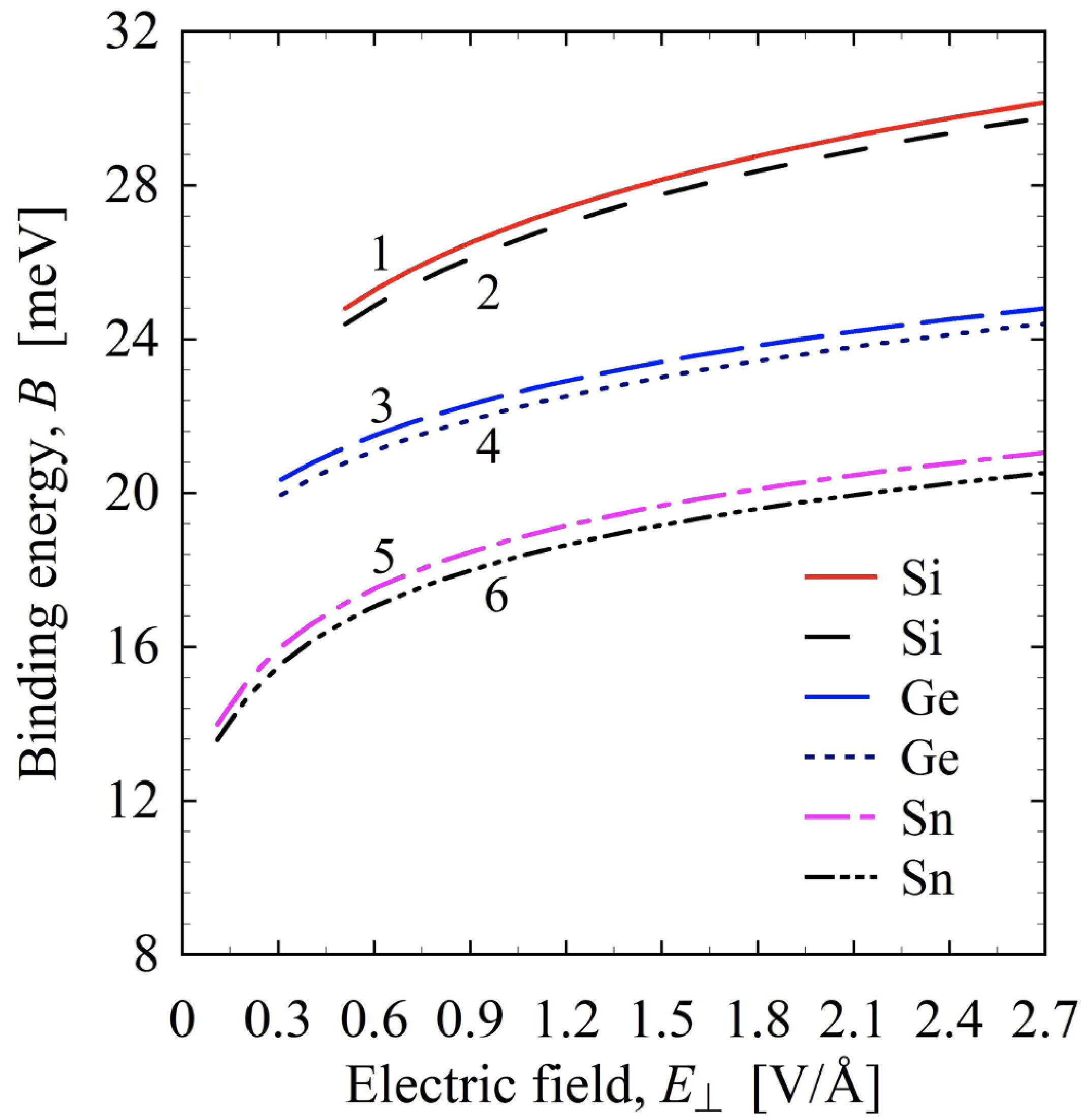}

\caption{(Color online) Dependencies of the BE of intravalley trions in freestanding silicene, germanene, and stanene on the applied electric field. Curves 1, 3, and 5 correspond to the trions formed by $A$ excitons coupling an electron ($X^{-}$) or hole ($X^{+}$), curves 2, 4, and 6 correspond to the trions formed by $B$ excitons coupling an electron ($X^{-}$) or hole ($X^{+}$). %Freestanding intravalley singlet state trions.
The plots for FS Xenes are truncated for the external electric field $E_{\perp}$ less than a critical field.}
\label{Freestanding}
\end{figure}

The input parameters for calculations of BEs of trions in the freestanding are given in Ref. \cite{KezSpir2022Xenes}. The formation of Wannier–Mott excitons due to the electron-hole interaction via the RK potential in semiconducting phases in Xenes monolayers occurs when the external electric field exceed some critical value which is unique to each material \cite{BBKPL2019}. A value of the critical electric field  is slightly different for $A$ excitons and for $B$ excitons. Following
\emph {ab initio} calculations \cite{Drummond2012} which determined that the crystal
structure of silicene becomes unstable around 2.6 V/\text{\AA}, we
consider in our calculations electric fields up to 2.7 V/\text{\AA}
and study the formations of trions in Xenes at the range of the external electric field from the critical value up to 2.7 V/\text{\AA}. Results of calculations of dependencies of BE of intravalley trions in a singlet state on the external electric field are presented in Fig. \ref{Freestanding}.

According to Fig. \ref{Freestanding}, the BE increases for all materials as $E_{\perp }$ increases. In addition, in FS Si, Ge, and Sn, we
observe a non-negligible difference in the BE of trions formed by
$A$ and $B$ excitons. These differences slightly increase as the electric field increases. The trion BEs for FS Ge and FS Sn are qualitatively
similar to FS Si, but they are smaller than freestanding silicene. The curves for FS Ge and Sn qualitatively resemble that of FS silicene, but at 2.7 V/\text{\AA} FS
germanene reaches a maximum trions BE of 24.8 (24.3) meV, and the
maximum BE for FS stanene is roughly 21.1 (20.5) meV,
significantly smaller than for FS silicene $-$ 30.1 (29.8) meV. In parentheses the BEs of trions formed by $B$ excitons are given. The percent differences
between the trion BE of FS Si and FS Ge and FS Si and FS Sn at the largest
electric field considered, are 82\% (81\%) and 70\% (69\%), respectively.

The following conclusions can be made: i. the increase of the BE as the external electric field increases; ii. BEs for FS silicene, germanene, and stanene are qualitatively
similar; iii. a non-negligible difference in the BE of trions formed by $A$ and $B$ excitons.

\begin{figure}[tbh]
\centering
\includegraphics[width=11.0cm]{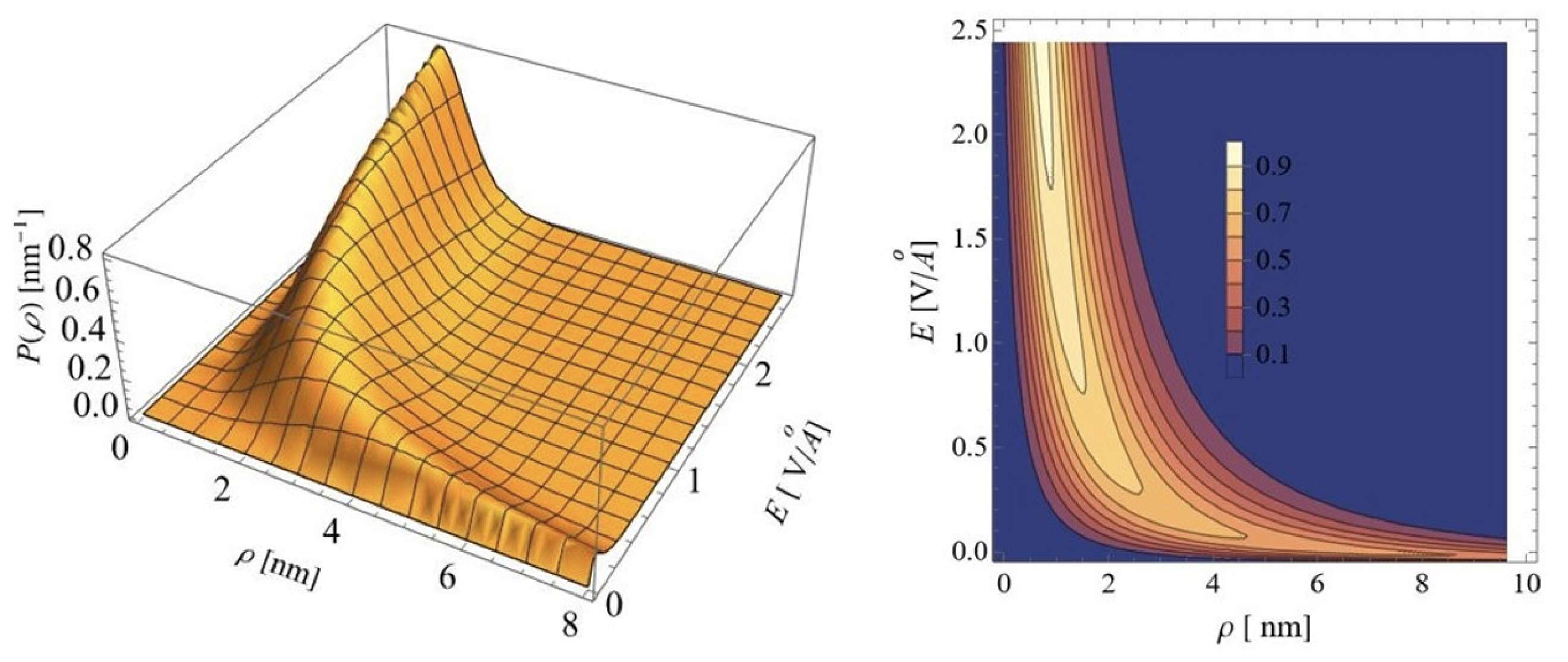}

\caption{(Color online) Dependence of the probability distribution of three particles in freestanding silicene
on hyperradius $\rho$ and applied electric field for intravalley $X^{-}$ trion.}
\label{Density}
\end{figure}

We calculated the probability distribution for
the intravalley trion for the spin-valley configuration shown in Fig. \ref{KvalleyTrion}$a$. In Fig. \ref{Density} the interparticle radial
probability distribution for intravalley trions in the silicene monolayer
is shown. The difference in the probability distribution is
related to the difference of the effective masses $\mu$ of intravalley trions. The analysis of the
dependence of the probability distribution of three particles
on the hyperradius $\rho$ and the external electric field leads to the following conclusion: the increase of the external electric field gives the increase of the trions BE and makes trions more compact since the greater binding energy
increases the trion formation probability.

\section{Conclusion}
\label{conc}
We predict the existence
of electrically controlled trions in Xenes monolayers.
We have applied the hyperspherical harmonics method to the calculation of
BEs for the trion, and we predict the formation of trions in freestanding Xenes when the external electric field perpendicular to monolayers is present. The results of BEs calculation for trions formed by $A$ and $B$ excitons show a non-negligible difference in trions energies that increases slightly as the electric field increases. The BEs of the intravalley trions can be tuned in the range of $24-31$  meV for silicene, $21-26$ meV for germanene, and  $14-20$ meV for stanene by varying the external electric field from the critical value that is specific for each material up to 2.7 V/\AA. Let us note that the trion binding energies in Xenes are of the same order as in TMDC monolayers \cite{Keztsik2024}. The dependence of the BE for silicene, germanene, and stanene as a function of the electric field is qualitatively similar. Our findings pave the way toward manipulating the trion BEs by an external electric field.
The results of calculations of the probability distribution show an increase in the compactness of trions with an increase of the electric field, since the greater binding energy
increases the trion formation probability.


\begin{thebibliography}{99}

\bibitem{Lampert58} M. A. Lampert, Mobile and immobile
effective-mass-particle complexes in nonmetallic solids, Phys. Rev. Lett.
\textbf{1}, 450 (1958).

\bibitem{Kormanyos} A. Korm\'{a}nyos, G. Burkard, M. Gmitra, J. Fabian, V. Z\'{o}lyomi, N.~D. Drummond, and V. Fal'ko, \textbf{k}$\cdotp$\textbf{p} theory for two-dimensional transition metal dichalcogenide semiconductors, 2D Mater. \textbf{2}, 022001
(2015).

\bibitem{RMP} G. Wang, A. Chernikov, M. M. Glazov, T. F. Heinz, X. Marie, T.
Amand, B. Urbaszek, Excitons in atomically thin transition metal
dichalcogenides, Rev. Mod. Phys. \textbf{90}, 21001 (2018).

\bibitem{MoS23Heinz} K. F. Mak, K. He, C. Lee, et al., Tightly bound trion
in monolayer MoS$_{2}$, Nat. Mater. \textbf{12}, 207 (2013).

\bibitem{NScRev2015} H. Yu, X. Cui, X. Xu, and W. Yao, Valley excitons in
two-dimensional semiconductors, Natl. Sci. Rev. \textbf{2}, 57 (2015).

\bibitem{BekReichman} T. C. Berkelbach and D. R. Reichman, Optical and
excitonic properties of atomically thin transition-metal dichalcogenides,
Annu. Rev. Condens. Matter Phys. 2018. \textbf{9}, 379--96 (2018).

\bibitem{Durnev2018} M. V. Durnev and M. M. Glazov, Excitons and trions in
two-dimensional semiconductors based on transition metal dichalcogenides,
Phys. Usp. \textbf{61}, 825 (2018).

\bibitem{Kezerashvili2019} R. Ya. Kezerashvili, Few-body systems in condensed matter physics, Few-Body Syst. \textbf{60}, 52 (2019).

\bibitem{Suris2022} M. A. Semina and R. A. Suris, Localized excitons and
trions in semiconductor nanosystems, Phys. Usp. \textbf{65}, 111 (2022).







%\bibitem{VargaPRB2016} D. W. Kidd, D. K. Zhang, and K. Varga, Binding energies and structures of two-dimensional excitonic complexes in transition metal dichalcogenides, Phys. Rev. B \textbf{93}, 125423 (2016).


\bibitem{KezFew2017} R. Ya. Kezerashvili and Sh. M. Tsiklauri, Trion and
biexciton in monolayer transition metal dichogenides, Few-Body Syst. \textbf{%
58}, 18 (2017).



\iffalse
\bibitem{FilikhinKez} I. Filikhin, R. Ya. Kezerashvili, Sh. M. Tsiklauri,
and B. Vlahovic, Trions in bulk and monolayer materials: Faddeev equations
and hyperspherical harmonics, Nanotechnology \textbf{29}, 124002 (2018).
\fi

\bibitem{FilikhinKezPL2018} I. Filikhin, R. Ya. Kezerashvili,
and B. Vlahovic, On binding energy of trions in bulk materials, Phys. Lett. A \textbf{382}, 787 (2018).
%\bibitem{Varga2020} J. Yan and K. Varga, Excited-state trions in two-dimensional materials, Phys. Rev. B \textbf{101}, 235435 (2020).

\bibitem{Frederico2023} K. Mohseni, M. R. Hadizadeh, T. Frederico, D. R. da
Costa, and A. J. Chaves, Trion clustering structure and BE in
two-dimensional semiconductor materials: Faddeev equations approach, Phys.
Rev. B \textbf{107}, 165427 (2023).

\bibitem{Keztsik2024}R. Ya. Kezerashvili, S. M. Tsiklauri, and A. Dublin, Trions in two-dimensional monolayers within the hyperspherical harmonics method:
application to transition metal dichalcogenides, Phys Rev. B \textbf{109}, 085406 (2024).

\bibitem{Molle2017} A. Molle, J. Goldberger, M. Houssa, Y. Xu, S.-C. Zhang, and D. Akinwande, Buckled
two-dimensional Xene sheets, Nat. Mater. \textbf{16}, 163 (2017).

\bibitem{Zheng2020} J. Zheng, Y. Xiang, C. Li, R. Yuan, F. Chi, and Y. Guo, All-optically controlled topological transistor based on Xenes, Phys. Rev. Appl. \textbf{14}, 034027 (2020).



\bibitem{Matthes2013a} L. Matthes, O. Pulci, and F. Bechstedt, Massive Dirac quasiparticles in the optical
absorbance of graphene, silicene, germanene, and tinene, J. Phys.: Condens. Matter
\textbf{25}, 395305 (2013).
%\bibitem{Grazianetti2020} C. Grazianetti, C. Martella, and A. Molle, Phys. Status Solidi RRL \textbf{14}, 1900439 (2020).

\bibitem{Acun2013} A. Acun, B. Poelsema, H. J. W. Zandvliet, and R. van Gastel, The instability of silicene on Ag(111), Appl. Phys. Lett. \textbf{103}, 263119 (2013).

\bibitem{Tao2015} L. Tao, E. Cinquanta, D. Chiappe, C. Grazianetti, M.
Fanciulli, M. Dubey, A. Molle, and D. Akinwande, Silicene field- effect transistors operating at room temperature, Nat. Nanotechnol. $\bm{10}$%
, 227 (2015).

\bibitem{Aufray2010} B. Aufray, A. Kara, S. Vizzini, H. Oughaddou, C. Léandri, B. Ealet, and G. Le Lay, Graphene-like silicon nanoribbons on Ag(110): A possible formation of silicene, Appl. Phys. Lett. \textbf{96}, 183102 (2010).

\bibitem{Padova2010} P. De Padova, C. Quaresima, C. Ottaviani, P. M. Sheverdyaeva, P. Moras, C. Carbone,
D. Topwal, B. Olivieri, A. Kara, H. Oughaddou, B. Aufray, and G. Le Lay, Evidence
of graphene-like electronic signature in silicene nanoribbons,  Appl. Phys. Lett. \textbf{96},
261905 (2010).


%\bibitem{Sante2019} D. Di Sante, X. Wu, M. Fink, W. Hanke, and R. Thomale, Phys. Rev. B \textbf{99}, 201106 (2019).

\bibitem{Drummond2012} N. D. Drummond, V. Z\'{o}lyomi, and V. I. Fal'ko, Electrically tunable band gap in
silicene, Phys. Rev. B \textbf{85},075423 (2012).

\bibitem{Ezawa2} M. Ezawa, Valley-polarized metals and quantum anomalous Hall effect in silicene, Phys. Rev. Lett. \textbf{109}, 055502 (2012).

\bibitem{Davila2014} M. E. Dávila, L. Xian, S. Cahangirov, A. Rubio, and G. L. Lay, Germanene: a novel
two-dimensional germanium allotrope akin to graphene and silicene,  New J. Phys
\textbf{16}, 095002 (2014).

\bibitem{Mannix2017} A. J. Mannix, B. Kiraly, M. C. Hersam, and N. P. Guisinger, Synthesis and chemistry
of elemental 2D materials, Nat. Rev. Chem. \textbf{1}, 0014 (2017).


\bibitem{Grazianetti2020} C. Grazianetti, C. Martella, and A. Molle, The Xenes generations: a taxonomy of
epitaxial single-element 2D materials, Phys. Status Solidi RRL \textbf{14}, 1900439 (2020).

%\bibitem{Drummond2012} N. D. Drummond, V. Z\'{o}lyomi, and V. I. Fal'ko, Phys. Rev. B\textbf{85},075423 (2012).


\bibitem{Sante2019} D. Di Sante, X. Wu, M. Fink, W. Hanke, and R. Thomale, Triplet superconductivity
in the Dirac semimetal germanene on a substrate, Phys. Rev. B \textbf{99}, 201106 (2019).

\bibitem{Li2013} L. Li, X. Wang, X. Zhao, and M. Zhao, Moiré superstructures of silicene on hexagonal boron nitride: A first-principles study, Phys. Lett. A $ \bm{377}$, 2628 (2013).

\bibitem{Khan2017} A. I. Khan, T. Chakraborty, N. Acharjee, and S. Subrina, Stanene-hexagonal boron
nitride heterobilayer: structure and characterization of electronic property, Sci. Rep. \textbf{7}, 16347 (2017).

\bibitem{Saxena2016} S. Saxena, R. P. Chaudhary, and S. Shukla, Stanene: atomically thick free-standing
layer of 2D hexagonal tin, Sci. Rep. \textbf{6}, 31073 (2016).

\bibitem{Ezawa2012sil} M. Ezawa, Quantum Hall effects in silicene,  J. Phys. Soc. Jpn. \textbf{81}, 064705 (2012).

\bibitem{KezSpir2022Xenes} R. Ya. Kezerashvili and A. Spiridonova, Effects of parallel electric and magnetic fields on Rydberg excitons in buckled
two-dimensional materials, Phys. Rev. B \textbf{103}, 165410 (2021).

\bibitem{Bechstedt2012} F. Bechstedt, L. Matthes, P. Gori, and O. Pulci, Infrared absorbance of silicene and
germanene, Appl. Phys. Lett. \textbf{100}, 261906 (2012).

\bibitem{Stille2012} L. Stille, C. J. Tabert, and E. J. Nicol, Optical signatures of the tunable band gap
and valley-spin coupling in silicene, Phys. Rev. B \textbf{86}, 195405 (2012).

\bibitem{Fadaie2016} M. Fadaie, N. Shahtahmassebi, and M. R. Roknabad, Effect of external electric field
on the electronic structure and optical properties of stanene,  Opt. Quantum Electron. \textbf{48}, 440 (2016).

\bibitem{Muoi2020} D. Muoi, N. N. Hieu, C. V. Nguyen, B. D. Hoi, H. V. Nguyen, N. D. Hien, N. A.
Poklonski, S. S. Kubakaddi, and H. V. Phuc, Magneto-optical absorption in silicene
and germanene induced by electric and Zeeman fields, Phys. Rev. B \textbf{101}, 205408 (2020).

\bibitem{Chowdhury2016} S. Chowdhury and D. Jana, A theoretical review on electronic, magnetic and optical
properties of silicene, Rep. Prog. Phys. \textbf{79}, 126501 (2016).

\bibitem{Zhai2018} X. Zhai, Y.-T. Wang, R. Wen, S.-X. Wang, Y. Tian, X. Zhou, W. Chen, and Z. Yang, Valley-locked thermospin effect in silicene and germanene with asymmetric magnetic field induced by ferromagnetic proximity effect, Phys. Rev. B \textbf{97}, 085410 (2018).

%\bibitem{EzawaAPS2012} M. Ezawa, Valley-polarized metals and quantum anomalous Hall effect in silicene, Phys. Rev. Lett. \textbf{109}, 055502 (2012).

\bibitem{Zhao2020} A. Zhao and B. Wang, Two-dimensional graphene-like Xenes as potential topological
materials, APL Mater. \textbf{8}, 030701 (2020).


\bibitem{Tsaran2014} V. Y. Tsaran and S. G. Sharapov, Landau levels and magnetic oscillations in gapped
Dirac materials with intrinsic Rashba interaction, Phys. Rev. B \textbf{90}, 205417 (2014).


\bibitem{Chen2018} C.-H. Chen, W.-W. Li, Y.-M. Chang, C.-Y. Lin, S.-H. Yang, Y. Xu, and Y.-F. Lin, Negative-differential-resistance devices achieved by band-structure engineering in
silicene under periodic potentials, Phys. Rev. Appl. \textbf{10}, 044047 (2018).

\bibitem{Lyu2019} J.-K. Lyu, S.-F. Zhang, C.-W. Zhang, and P.-J.Wang, Stanene: a promising material for new electronic and spintronic applications, Ann. Phys. (Berlin, Ger.) \textbf{531},
1900017 (2019).

\bibitem{Glavin2020} N. R. Glavin, R. Rao, V. Varshney, E. Bianco, A. Apte, A. Roy, E. Ringe, and P. M.
Ajayan, Emerging applications of elemental 2D materials, Adv. Mater. \textbf{32}, 1904302
(2020).

\bibitem{Tao2020} C. Grazianetti, C. Martella, and A. Molle, 8 - Two-dimensional Xenes and their device
concepts for future micro- and nanoelectronics and energy applications, edited by L.
Tao and D. Akinwande, Micro and Nano Technologies (Elsevier, 2020), pp. 181-219.


\bibitem{Bechstedt2021} F. Bechstedt, P. Gori, and O. Pulci, Beyond graphene:
clean, hydrogenated and halogenated silicene, germanene,
stanene, and plumbene, Prog. Surf. Sci.
\textbf{96}, 100615 (2021).

\bibitem{Liu2011} C.-C. Liu, W. Feng, and Y. Yao, Quantum spin Hall effect
in silicene and two-dimensional germanium, Phys. Rev.
Lett. \textbf{107}, 076802 (2011).

\bibitem{Ezawa2015} M. Ezawa, Monolayer topological insulators: silicene, germanene, and stanene, J. Phys. Soc. Japan \textbf{84}, 121003 (2015).

\bibitem{Matthes2016} L Matthes, S. K\"ufner, J. Furthm\"uller, and F. Bechstedt, Quantization and topological states in the spin Hall conductivity
of low-dimensional systems: An ab initio study,
Phys. Rev. B \textbf{93}, 121106 (2016).

\bibitem{Matusalem2017} F. Matusalem, D. S. Koda, F. Bechstedt, M. Marques,
and L. K. Teles, Deposition of topological silicene, germanene
and stanene on graphene-covered sic substrates,
Sci. Rep. \textbf{7}, 15700 (2017).

\bibitem{Yu2017} X.-L. Yu, L. Huang, and J. Wu, From a normal insulator
to a topological insulator in plumbene, Phys. Rev. B \textbf{95},
125113 (2017).



\bibitem{BBKPL2019} M. N. Brunetti, O. L. Berman, and R. Ya. Kezerashvili, Can freestanding Xene monolayers behave as excitonic insulators?, Phys. Lett. A \textbf{383}, 482 (2019).

\bibitem{BBKPRB2019} M. N. Brunetti, O. L. Berman, and R. Ya. Kezerashvili, Optical properties of excitons in buckled two-dimensional materials in an external electric field, Phys. Rev. B \textbf{98}, 125406 (2018).


\bibitem{Pulci2023} O. Pulci, P. Gori, D. Grassano, M. D’Alessandro, and F. Bechstedt, Transitions in Xenes between excitonic, topological and trivial insulator phases: influence of screening, band dispersion and external electric field, SciPost Phys. \textbf{15}, 025 (2023).


\bibitem{Ritova} N. S. Rytova, Screened potential of a point charge in a thin film, Proc. MSU Phys., Astron. \textbf{3}, 30
(1967),
https://www.researchgate.net/publication/320224883\_Screened\_potential\_of%
\_a\_point\_charge\_in\_a\_thin\_film.

\bibitem{Keldysh} L. V. Keldysh, Coulomb interaction in thin semiconductor
and semimetal films, JETP Lett. \textbf{29}, 658 (1979).


\bibitem{Tabert} C. J. Tabert and E. J. Nicol, Dynamical polarization function, plasmons, and screening in silicene and other buckled honeycomb lattices, Phys. Rev. B \textbf{89}, 195410 (2014).

\iffalse
\bibitem{CastroNeto2009a} A. H. Castro Neto, F. Guinea, N. M. R. Peres, K.
S. Novoselov, and A. K. Geim, The electronic properties of graphene, Rev. Mod. Phys. 81, 109 (2009).

\bibitem{Abergel2010} D. S. L. Abergel, V. Apalkov, J. Berashevich, K.
Ziegler, and T. Chakraborty, Properties
of graphene: a theoretical perspective, Adv. Phys. 59, 261 (2010).

\bibitem{Ezawa1} M. Ezawa, A topological insulator and helical zero mode in silicene under an inhomogeneous
electric field, New J. Phys. \textbf{14}, 033003 (2012).

\bibitem{Ezawa3} M. Ezawa, Spin-valley optical selection rule and strong circular dichroism in silicene, Phys. Rev. B \textbf{86}, 161407(R) (2012).
\fi
\bibitem{Berkelbach2013} T. C. Berkelbach, M. S. Hybertsen, and D. R.
Reichman, Theory of neutral and
charged excitons in monolayer transition metal dichalcogenides, Phys. Rev. B \textbf{88}, 045318 (2013).

\bibitem{Fogler2014}  M. Fogler, L. Butov, and K. Novoselov, High-temperature superfluidity with indirect
excitons in van der Waals heterostructures, Nat. Commun. \textbf{5}, 4555 (2014).

\bibitem{Berman2016} O. L. Berman and R. Ya. Kezerashvili, High-temperature superfluidity of the two component Bose gas in a transition metal dichalcogenide bilayer, Phys. Rev. B \textbf{93},
245410 (2016).

\bibitem{Wang2018} G. Wang, A. Chernikov, M. M. Glazov, T. F. Heinz, X. Marie, T. Amand, and B.
Urbaszek, Colloquium: excitons in atomically thin transition metal dichalcogenides, Rev. Mod. Phys. \textbf{90}, 021001 (2018).

\bibitem{Berman2017} O. L. Berman and R. Ya. Kezerashvili, Superfluidity of dipolar excitons in a transition
metal dichalcogenide double layer, Phys. Rev. B \textbf{96}, 094502 (2017).

\bibitem{Rubio} P. Cudazzo, I.~V. Tokatly, and A. Rubio, Dielectric screening in two-dimensional
insulators: implications for excitonic and impurity states in graphane,  \prb {\bf 84},
085406 (2011).


\bibitem{Avery} J. Avery, Hyperspherical Harmonics, Kluwer Academic,
Dordrecht, 1989.

\bibitem{JibutiSh} R.I. Jibuti and K. V. Shitikova: Method of
hyperspherical functions in atomic and nuclear physics, Energoatomizdat,
Moscow, 270p. 1993. (in Russian).

\bibitem{FM} L.D. Faddeev and S.P. Merkuriev, Quantum scattering theory for
several particle systems (Kluwer Academic, Dordrecht, 1993) pp. 398.

\bibitem{FilKez2024}I. Filikhin, R. Ya. Kezerashvili, and B. Vlahovic, The charge
and mass symmetry breaking in the KK$\bar{\text{K}}$ system, J. Phys. G: Nucl. Part. Phys. \textbf{51}, 035102 (2024).

\iffalse
\bibitem{Balendhran} S. Balendhran, S.Walia, H. Nili, S. Sriram, and M.
Bhaskaran, Elemental analogues of graphene: silicene, germanene, stanene, and phosphorene, Small $\bm{11}$, 640 (2015).

\bibitem{Ni2012} Z. Ni, Q. Liu, K. Tang, J. Zheng, J. Zhou, R. Qin, Z. Gao,
D. Yu, and J. Lu, Tunable bandgap in silicene and germanene, Nano Lett. $\bm{12}$, 113 (2012).


\bibitem{Tanatar} B. Tanatar and D.~M. Ceperley, Ground state of the two-dimensional electron gas, \prb {\bf 39}, 5005
(1989).
\bibitem{Wigner1934}E. Wigner, On the interaction of electrons in metals, Phys. Rev. \textbf{46}, 1002 (1934).

\bibitem{Andrei1988}E. Andrei, G. Deville, D. Glattli, F. Williams, E. Paris, and B. Etienne, Observation of a magnetically induced Wigner solid, Phys. Rev. Lett. \textbf{60}, 2765 (1988)

\bibitem{Goldman1990}V. Goldman, M. Santos, M. Shayegan, J. Cunningham, Evidence
for two-dimentional quantum Wigner crystal, Phys. Rev. Lett. \textbf{65}, 2189 (1990).

\bibitem{Williams1991}F. Williams, P. Wright, R. Clark, E. Andrei, G. Deville, D. Glattli, et al. Conduction
threshold and pinning frequency of magnetically induced Wigner solid,
Phys. Rev. Lett. \textbf{66}, 3285 (1991).

\bibitem{Smole2021}T. Smol\'{e}nski, P. E. Dolgirev, C. Kuhlenkamp, A. Popert,
Y. Shimazaki, P. Back, et al., Signatures of Wigner crystal of electrons in a monolayer semiconductor, Nature \textbf{595}, 53 (2021).
\bibitem{Sung2023} J. Sung, J. Wang, I. Esterlis, P. A. Volkov, G. Scuri, Y. Zhou, et al.,
Observation of an electronic microemulsion
phase emerging from a quantum crystal-to-liquid transition, (2023), arXiv:2311.18069 [cond-mat.str-el].

\bibitem {Xiang2024} Z. Xiang, H. Li, J. Xiao, M. H. Naik, Z. Ge, Z. He, et al.,
Quantum melting of a disordered wigner solid, (2024), arXiv:2402.05456 [cond-mat.str-el].

\bibitem {Zhou2021}Y. Zhou, J. Sung, E. Brutschea, I. Esterlis, Y. Wang, G. Scuri, et al.,
Bilayer Wigner
crystals in a transition metal dichalcogenide heterostructure. Nature
\textbf{595}, 48 (2021).

\bibitem {Huang2021}X. Huang, T. Wang, S. Miao, C. Wang, Z. Li, Z. Lian, et al., Correlated
insulating states at fractional fillings of the WS$_{2}$/WSe$_{2}$ moiré lattice,
Nat. Phys. \textbf{17}, 715(2021).

\bibitem {Miao2021}S. Miao, T. Wang, X. Huang, D. Chen, Z. Lian, C. Wang, M. Blei, et al., Strong interaction
between interlayer excitons and correlated electrons in WSe$_{2}$/WS$_{2}$
moiré superlattice, Nat. Commun. \textbf{12}, 1 (2021).

\bibitem{Shapir2019}I. Shapir, A. Hamo, S. Pecker, C. P. Moca, \"{O}Legeza, G. Zarand, and S. Ilani, Imaging the electronic Wigner crystal in one dimension, Science, \textbf{364}, 870 (2019).


\bibitem{Swierkowski1991}L. Swierkowski, D. Neilson, and J. Szym\'{a}nski, Enhancement of Wigner crystallization in multiple-quantum-well
structures, Phys. Rev. Lett. \textbf{67}, 240 (1991).

\bibitem{Rapisarda1996} F. Rapisarda and G. Senatore, Diffusion Monte Carlo study of
electrons in two-dimensional layers, Aust. J. Phys. \textbf{49}, 161 (1996).


\bibitem{Goldoni1997}G. Goldoni and F.M. Peeters, Wigner crystallization in quantum electron bilayers, Europhys. Lett. \textbf{37}, 293
(1997); ibid. \textbf{38}, 319 (1997).

\bibitem {Attaccalite2002} C. Attaccalite, S. Moroni, P. Gori-Giorgi, and G. B. Bachelet. Correlation energy and spin polarization in the 2D electron gas, Phys. Rev. Lett. \textbf{88}, 256601 (2002).

\bibitem {Noda2002} Y. Noda and M. Imada, Quantum phase transitions to charge-ordered and Wigner-crystal states under the interplay of lattice commensurability and long-range Coulomb interactions,
Phys. Rev. Lett. \textbf{89}, 176803 (2002).
\bibitem {Valenzuela2003}B. Valenzuela, S. Fratini and D. Baeriswyl, Charge and
spin order in one-dimensional electron systems with long-range Coulomb interactions, Phys. Rev. B \textbf{68}, 045112 (2003).

\bibitem{Drummond2009}N. D. Drummond and R. J. Needs, Phase diagram of the low-density two-dimensional homogeneous electron gas, Phys. Rev. Lett. \textbf{102}, 126402 (2009).

\bibitem {Siegmund2009} M. Siegmund, M. Hofmann and O. Pankratov, Density functional
study of collective electron localization: detection by persistent
current, J. Phys.: Condens. Matter \textbf{21} 155602 (2009).

\bibitem{Zarenia2017}M. Zarenia, D. Neilson, B. Partoens, and F. M. Peeters, Wigner crystallization in transition metal dichalcogenides: a new approach to correlation energy, Phys. Rev. B \textbf{95}, 115438 (2017).

\bibitem{Germani2020}J. Kn\"{o}rzer, M. J. A. Schuetz, G. Giedke, et al., Wigner crystals in two-dimensional transition-metal dichalcogenides: spin physics and readout, Phys. Rev. B \textbf{101}, 125101 (2020).

\bibitem{Padhi2021}B. Padhi, R. Chitra, P. W. Phillips, Generalized Wigner crystallization in moiré materials, Phys. Rev. B \textbf{103}, 125146 (2021).

\bibitem{Brem2022}S. Brem and E. Malic, Terahertz fingerprint of monolayer Wigner crystals, Nano Lett. \textbf{22}, 1311 (2022).

\bibitem{Das Sarma2024}Y. Huang and S. Das Sarma, Electronic transport, metal-insulator transition, and Wigner crystallization in transition metal dichalcogenide monolayers, arXiv:2404.03488v1 [cond-mat.mes-hall]  (2024).

\bibitem{RKCrystal2011} O. L. Berman, R. Ya. Kezerashvili, Restricted three body problem in semiconductor heterostructure, Few-Body Syst. \textbf{50}, 407 (2011).

\bibitem{RKCrystal2014} O. L. Berman, R. Ya. Kezerashvili, and S. M. Tsiklauri, Trions in coupled quantum wells and Wigner crystallization, Int. J. Mod. Phys. B \textbf{28}, 1450064 (2014).


\bibitem{RKCrystal2022}I. V. Bondarev, O. L. Berman, R. Ya. Kezerashvili, and Yu. E. Lozovik, Crystal phases of charged interlayer excitons in van der Waals heterostructures, Comm. Phys. \textbf{4}, 134 (2021).

\bibitem{BondarevCrystal2022}I. V. Bondarev and Yu. E. Lozovik, Magnetic-field-induced Wigner crystallization of
charged interlayer excitons in van der Waals
heterostructures, Comm. Phys. \textbf{5}, 315 (2022).

\bibitem{RKVK2022} R. Ya. Kezerashvili and V. Ya. Kezerashvili, Charge-dipole and dipole-dipole interactions in two-dimensional materials, Phys. Rev. B \textbf{105}, 205416 (2022).


\bibitem{Platzman}P. M. Platzman and H. Fukuyama, Phase diagram of the two-dimensional electron liquid, Phys. Rev. B \textbf{10}, 3150 (1974).
\fi

\end{thebibliography}
\end{document}